\documentclass{article}
\usepackage{kr}

\usepackage{times}
\usepackage{soul}
\usepackage{url}
\usepackage[hidelinks]{hyperref}
\usepackage[utf8]{inputenc}
\usepackage[small]{caption}
\usepackage{graphicx}
\usepackage{amsmath}
\usepackage{amsthm}
\usepackage{booktabs}
\usepackage{algorithm}
\usepackage{algorithmic}
\usepackage{xcolor}
\usepackage{subcaption}
\usepackage{dirtytalk}

\title{Automated Reasoning in Systems Biology: a Necessity for Precision Medicine}

\author{%
Pedro Zuidberg Dos Martires$^1$\and
Vincent Derkinderen$^{2}$\and
Luc De Raedt$^{1,2}$\and
Marcus Krantz$^3$ \\
\affiliations
$^1$Center for Applied Autonomous Systems, Örebro University, Sweden\\
$^2$Department of Computer Science, KU Leuven, Belgium\\
$^3$School of Medical Sciences, Faculty of Medicine and Health, Örebro University, Sweden
\emails
pedro.zuidberg-dos-martires@oru.se
}

\usepackage{xspace}
\newcommand{\cf}{cf.\xspace}
\newcommand{\eg}{e.g.\xspace}

\newcommand{\sysbio}{SysBio\xspace}

\begin{document}

\maketitle

\begin{abstract}

Recent developments in AI have reinvigorated pursuits to advance the (life) sciences using AI techniques, thereby creating a renewed opportunity to bridge different fields and find synergies.
Headlines for AI and the life sciences have been dominated by data-driven techniques, for instance, to solve protein folding with next to no expert knowledge.
In contrast to this, we argue for the necessity of a formal representation of expert knowledge -- either to develop explicit scientific theories or to compensate for the lack of data.
Specifically, we argue that the fields of knowledge representation (KR) and systems biology (SysBio) exhibit important overlaps that have been largely ignored so far. This, in turn, means that relevant scientific questions are ready to be answered using the right domain knowledge (SysBio), encoded in the right way (SysBio/KR), and by combining it with modern automated reasoning tools (KR). Hence, the formal representation of domain knowledge is a natural meeting place for SysBio and KR.
On the one hand, we argue that such an interdisciplinary approach will advance the field SysBio by exposing it to industrial-grade reasoning tools and thereby allowing novel scientific questions to be tackled.
On the other hand, we see ample opportunities to move the state-of-the-art in KR by tailoring KR methods to the field of SysBio, which comes with challenging problem characteristics, \eg scale, partial knowledge, noise, or sub-symbolic data.
We stipulate that this proposed interdisciplinary research is necessary to attain a prominent long-term goal in the health sciences: precision medicine.



\end{abstract}

\section{Introduction}

Precision medicine, as a research endeavour, has made the promise to deliver individualized therapies to patients by considering individuals' medical history and genetic make-up.
The idea is to create a faithful digital model for individual patients and how they would react to specific drug treatments.
While data-driven techniques, for instance in protein folding~\cite{Abramson2024}, are contributing to this endeavour and are part of the solution, we argue for the necessity of a formal representation of expert knowledge -- either to develop explicit scientific theories or to compensate for the lack of data.

To attain the goals set out by precision medicine, we require an understanding of how system level function emerges from the interplay between molecular mechanisms, as genetic variation and drugs exert their effects on the molecular level, but health and disease are system level properties.
Unfortunately, progress in this area has been rather slow. On the one hand, we attribute this to the inherent hardness of the research endeavour itself. On the other hand, we also see a problem with the techniques being deployed -- specifically in the language used to represent and reason over bio-medical knowledge.


\citeauthor{lazebnik2002can} describes this problem eloquently in an entertaining thought-experiment, where he applies a biologist's research methodolgy to repairing a broken transistor radio~\cite{lazebnik2002can}.
The bottom line of this thought experiment is that biologists may be good at cataloguing and characterising the radio's components and their connections, but would struggle to understand the system as a whole, and hence struggle to use their knowledge to repair the radio. The problem \citeauthor{lazebnik2002can} identified is the language in which biologists tend to represent knowledge: the graphical representation that is used to describe components of biological systems lacked formal semantics. This is in stark contrast to schematics used by engineers.
\citeauthor{lazebnik2002can} saw this also as the root cause of a paradox noted by David Papermaster: ``the more facts we learn, the less we understand the process we study''.


Since \citeauthor{lazebnik2002can} raised his concerns, tremendous progress has been made towards formalising biological knowledge. For instance, with the advent of the Systems Biological Graphical Notation (SBGN)~\cite{SBGN2009}.
However, we argue that the techniques currently used in \sysbio are not fit to tackling the problem of precision medicine.
At the core we identify combinatorial spaces as the pain-point of current \sysbio languages as this often leads to a mismatch between model resolution and data resolution. That is, model variables do not correspond to system observables. 
We describe this issue further in Section~\ref{sec:sota}. 
In addition to the presence of combinatorial complexity, \sysbio has challenges very related to KR: dealing with noisy and partial knowledge (of different resolution levels), as well as temporal data.

We view \textbf{KR techniques as a viable solution for these problems, and even argue that their integration into the \sysbio toolbox is necessary to advance the field of precision medicine}. We emphasize that this effort must, however, be carried by specialists from both fields to ensure proper tailoring to the expert knowledge present in \sysbio. 

We specifically consider logic to be a natural meeting ground between both fields, and envision the expert knowledge present in \sysbio to be modelled using it, albeit with probabilistic and neural extensions~\cite{Derkinderen2024semirings,MarraDMR24} to help deal with noise and sub-symbolic data. Furthermore, we consider efficient learning and reasoning algorithms to be key for incorporating sensory data (\textit{in vivo} measurements) and the efficient analysis of developed models.

We provide a brief primer on systems biology in Section~\ref{sec:primer}, emphasizing the need to create a system level understanding from lower level observations (which has consequences for whether a knowledge representation is appropriate).
Section~\ref{sec:combinatorial} explains our position by hinting how a signal transduction network can be modelled using logic formulas, and what reasoning tasks are of interest. 
We also explain how these networks may lead to combinatorial complexity challenges, which has impacted existing formalisms (cf. Section~\ref{sec:sota}).
Finally, we elaborate on logic as a meeting ground, in section~\ref{sec:marriage}. 

\section{Systems Biology in a Nutshell}\label{sec:primer}


Systems biology is an approach to understand how the function of a biological system emerges from the interplay between its parts~\cite{Eberhard2022}. To reach this goal, SysBio develops computational models, that allow for computer assisted reasoning on scientific hypotheses. The first and arguably most important step in this process is the formalisation of expert knowledge into a language that can be used by an automatic reasoner. Hence, the overlap with KR is striking, as is the lack of significant synergies between the fields.

(Molecular and cellular) biology is traditionally dominated by the reductionist approach: To study these immensely complex systems, they are broken down in minimal functional parts that are - as far as possible - studied in isolation. This has led to a high density of local knowledge, often centered on a specific cellular component or process. However, as discussed above, the field has struggled to turn this into a corresponding understanding at the system level due to the complexity of those systems as a whole, and the general lack of a common formal language~\cite{Bender2021,lazebnik2002can}. 
From a health care perspective, interest is on the system level properties, such as health and disease, while both patient differences (genetics) and treatments (drugs) act at the molecular level. Hence, bridging these levels by turning molecular level domain knowledge into a system level understanding is the key challenge in SysBio and a central pillar of future precision medicine. 


The holy grail of SysBio is to create a whole cell model (and ultimately a whole patient model), which accounts for the molecular function of all cellular components. This goal has until now been achieved only once, for a very simple bacterium ~\cite{Karr2012}. While this is an impressive feat, the corresponding tools to understand human cells are far away. In particular, this bacterium lacks the sophisticated signal transduction network, which appears to be the most challenging aspect to model in human cells. We describe this in more detail in the next section.

\section{\sysbio and Combinatorial Spaces}\label{sec:combinatorial}

The term cellular signal transduction network (STN) denotes the complex series of molecular events that occur within a cell in response to external and internal signals. These networks are crucial for cells to communicate with their environment and respond appropriately to various stimuli such as hormones, growth factors, nutrients, stressors, or in the case of precision medicine therapeutics drug treatments.

\begin{figure}
    \centering
    \includegraphics[width=\linewidth]{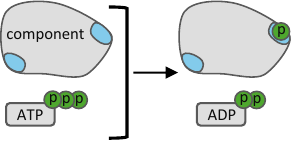}
    \caption{Illustration of a site-specific state change:
    a phosphor from a molecule (ATP) is attached to a specific position (also called site) of a protein (also called component). The depicted component has two sites indicated by the blue color -- $S_0$ (bottom left) and $S_1$ (top right).
    }
    \label{fig:state-change}
\end{figure}

The process typically involves a signaling molecule binding to a receptor on the cell surface or inside the cell, triggering a cascade of molecular events. These events often involve the activation of proteins called kinases, which can modify other proteins by adding phosphate groups to them, thereby altering their activity or localization within the cell. We illustrate this in Figure~\ref{fig:state-change}.

Such series of events ultimately lead to changes in gene expression, cell metabolism, cell proliferation, differentiation, or other cellular responses.
In turn, dysregulation of signal transduction networks can contribute to various diseases, including cancer, diabetes, and neurological disorders.
STNs are hence essential for coordinating various cellular processes and maintaining cellular homeostasis.

One of the goals of systems biology is now to express STNs using formal languages.
For instance, we can write down the reaction depicted in Figure~\ref{fig:state-change} using propositional logic formulas as follows: 
\begin{align*}
R^{t+1} &\leftrightarrow \neg S_1^t \land  ATP^t
\\
S_1^{t+1} &\leftrightarrow S_1^{t} \lor  R^t
\end{align*}
In this formalism $S_1$ denotes site $1$ of the component (upper right site in Figure~\ref{fig:state-change}) and $R$ denotes the reaction.
The superscripts indicate the time step\footnote{Using temporal logic formulas to represent reactions and sites is inspired by the work of~\cite{Romers2020}.}.
We can read the first line as \say{the reaction $R$ will happen in the next time step if and only if the site $S_1$ is not phosphorylated in the current time step and if an ATP molecule is present.}
The second line simply states that \say{the site $S_1$ is phosphorylated in the next time step if and only if it is already phosporylated in the previous time step or if the reaction $R$ happens.} Similar temporal formulas can be written down for the ATP and ADP molecules, and the other site of the component. 

\begin{figure}
    \centering
    \includegraphics[width=\linewidth]{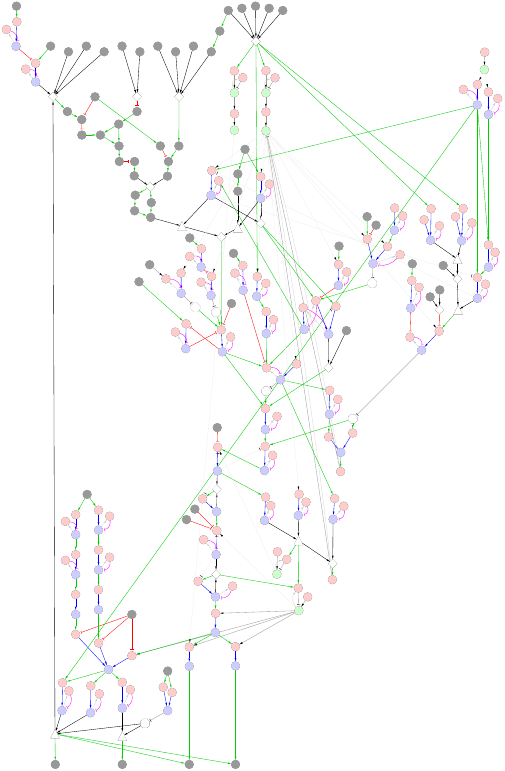}
    \caption[fragile]{The NLRP3 inflammasome signalling pathway constitutes a tiny part of the human STN and is a key regulator of inflammation processes in human cells. 
    We depict in mechanistic detail using a reaction-contingency-based formalism (\cf Section \ref{sec:sota}) the assembly and activation of the inflammasome. The figure is reproduced from \cite{krantz2023detailed}.}
    \label{fig:inflamma}
\end{figure}

STNs do not involve a single such reaction but multiple cascading reactions that are highly interconnected, with multiple signaling pathways often converging and interacting with each other, and components often having many modification sites. This gives rise to the highly combinatorial nature of STNs leading to components having an exponential number of possible configurations.
We give in Figure~\ref{fig:inflamma} a graphical representation of a fragment of such a signaling pathway -- an STN is the union of multiple of these pathways.

However, in reality not all configurations do indeed occur in living cells~\cite{Hlavacek2003}. This means that, \textbf{even though the space is highly combinatorial, it also exhibits a high degree of structure. }
Therefore, tools from the field of KR are perfectly suited for aiding in the modelling of STNs (\cf Section~\ref{sec:marriage}), instead of using tools already present in the \sysbio literature (\cf Section~\ref{sec:sota}).

\begin{figure}
    \centering
    \includegraphics[width=\linewidth]{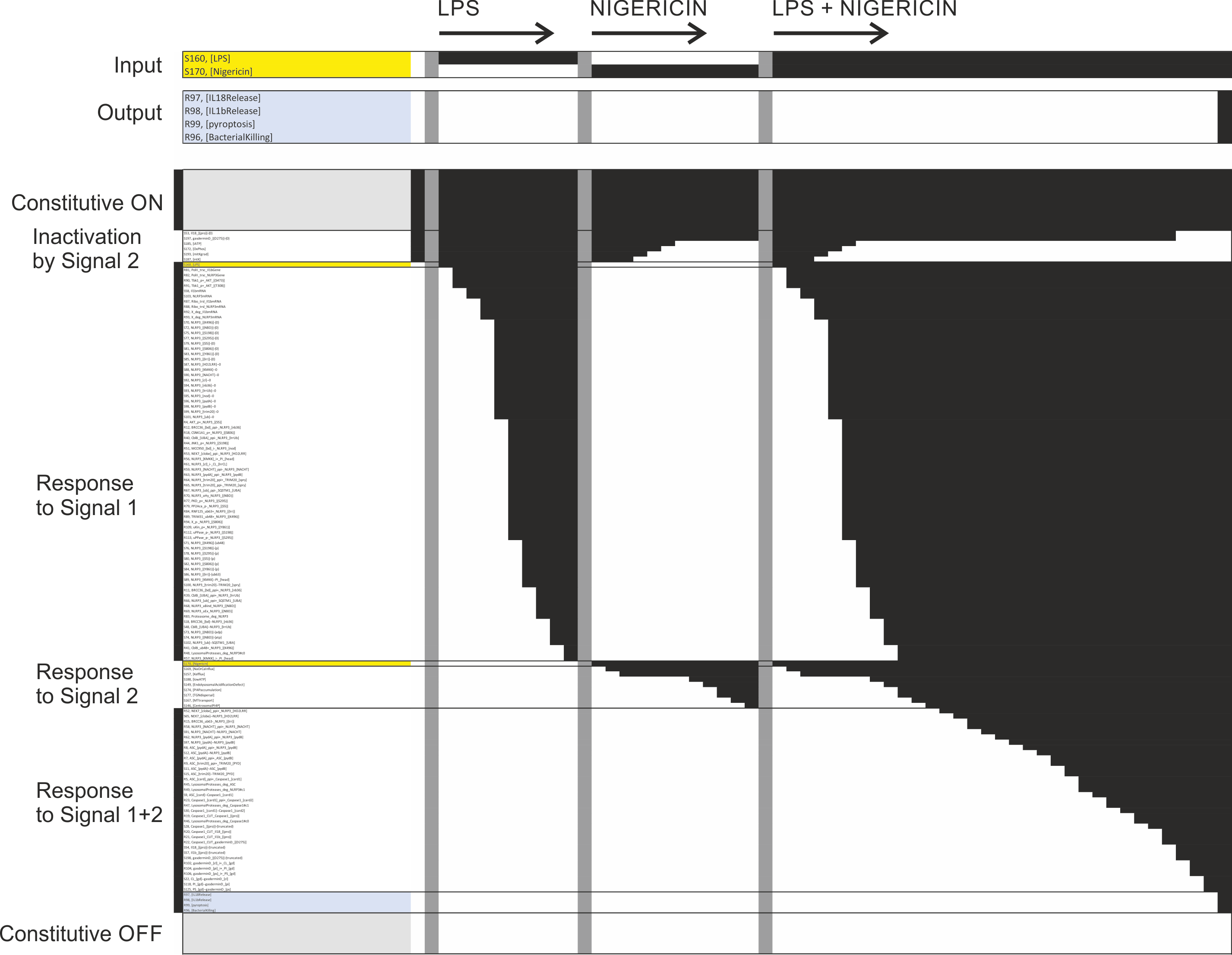}
    \caption[fragile]{Bipartite Boolean simulation of the NLRP3 inflammasome pathway from Figure~\ref{fig:inflamma}.
    Each row represent the value of a Boolean model variable, \eg modifications at component sites and reactions, over time, where black corresponds to true and white to false.
    The three grey columns indicate interventions in the system where at specific time points a subset of the state variables were set by hand.
    Each intervention simulates exposure of the system to different signals that together (but not individually) trigger inflammasome activation.
    The simulation shows that only after the third intervention (third greyed-in column) the initial conditions are set such that the pathway is active (indicated by the state variables in the blue boxes turning true).
    For further details we refer the reader to \cite{krantz2023detailed}, from where we reproduced the figure.}
    \label{fig:attractor_state}
\end{figure}

Once a formal model of a system has been obtained, one can start querying it. A simple query would be to simply simulate the system and study its behavior over time.
For instance, using rxncon, the reaction-contingency language ~\cite{Romers2017}, the knowledge base behind the regulatory graph in Figure~\ref{fig:inflamma} can be translated to temporal Boolean formulas ~\cite{Romers2020}. 
In Figure~\ref{fig:attractor_state} we illustrate the simulation of the pathway from its natural off-state.



\section{Problems with Current Techniques}\label{sec:sota}

\begin{figure*}
\centering
    \includegraphics[width=0.85\linewidth]{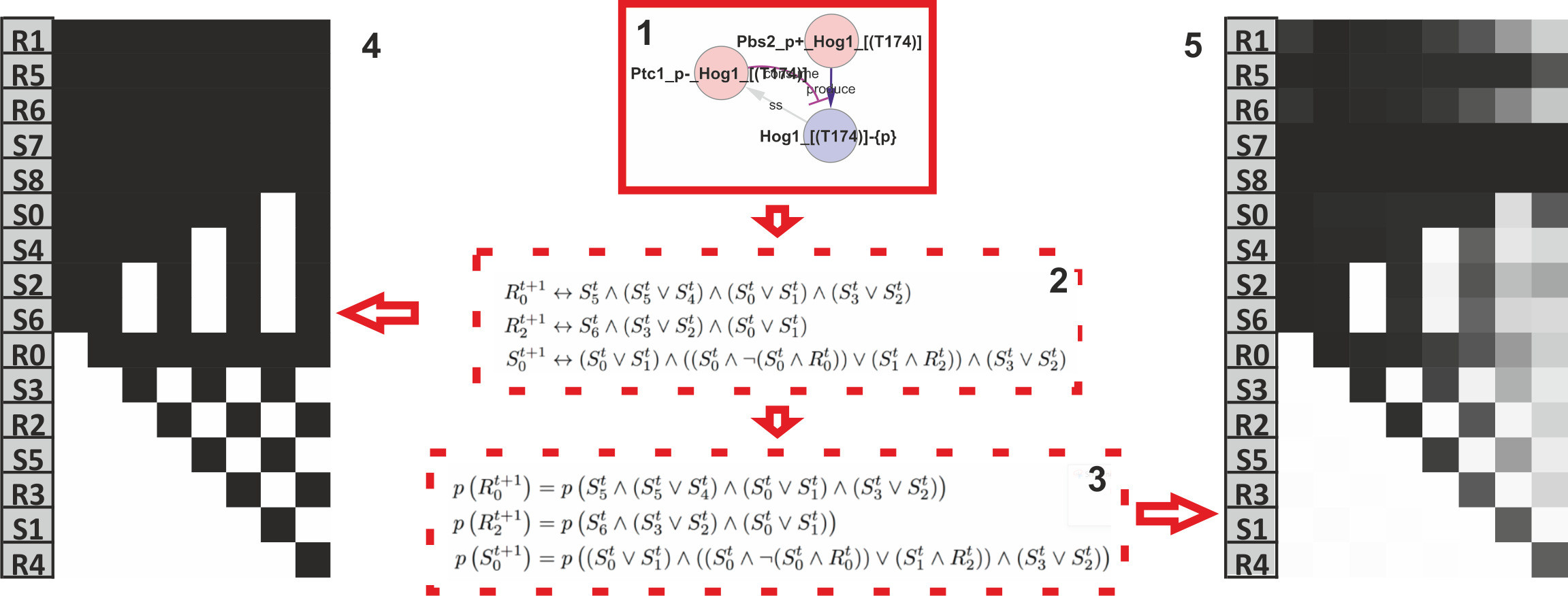}
    \caption[fragile]{In the middle at the top (Panel 1) we have a fragment of a reaction-contingency network with two reactions (red nodes) and one site-specific state (blue node). Using the bipartite Boolean network language~\cite{Romers2020} we can express these as temporal rules in propositional logic (Panel 2). These rules can then be used in a straightforward fashion to simulate the system over time by providing initial conditions and recursively computing the left-hand side of the equivalences. We can see such a simulation in Panel~4, where we give the trajectories not only for the fragment in Panel~1 but for the the entire set of variables involved in the pathway. Alternatively, we can use probabilistic logics (Panel 3). 
    Simulating the pathway using these probabilistic transition rules now leads to simulations of the pathway where site and reaction variables are not deterministically true or false anymore. This is indicated by the greyed-in cells in Panel~5.
    Using these probabilities one can now perform a quantitative analysis of the system over time -- a feature that is not possible using non-probabilistic representations. 
    }
    \label{fig:rcgraph2poolnet}
\end{figure*}

Current strategies to model STNs can be divided into three general formalism types depending on how they represent component states: 1) microstate based formalism, 2) component level formalism, and 3) reaction-contingency based formalisms. Incidentially, they correspond to the three SBGN graphical languages \cite{SBGN2009}.
We describe below the key characteristics of these methods and argue why \textbf{reaction-contingency based methods are the only viable option for precision medicine}.

\textbf{1)} The most commonly used modelling techniques, as of date, fall into the class of microstate based formalisms.  
They build on the successful methods of metabolic modelling and are easily represented mathematically by ordinary differential equations (ODEs), which give access to an extensive toolbox. They have been successfully used to build and analyse small-scale signal transduction models.

However, a major drawback of microstate based formalisms is their need to explicitly enumerate all (relevant) microstates, i.e. all combinations of site-specific states (\cf Figure~\ref{fig:state-change}).
For larger signaling pathways, \eg the human inflammosome in Figure~\ref{fig:inflamma} or even the entire STN, this constitutes a fundamental barrier that cannot be overcome. 
Furthermore, the formalism exhibits a low degree of composability,
which explains why large-scale models are divided into small, independent, and separated modules (\cf \url{reactome.org}).
While these formalisms are useful when studying small specific sub-parts of an STN, this approach is not viable at the scale required for precision medicine.




\textbf{2)} 
Component level formalisms take the opposite approach and only represent components (not their states).
While this yields an approach that easily scales and makes the network amenable to computer-assisted reasoning~\cite{Nilsson2022artificial}, critical information is discarded.
By abstracting away the site specific states, component level formalisms discard critical knowledge on how information is encoded and processed in the cell. Importantly, components may have different active states with different target specificities, as illustrated by Cdk1, the main kinase driving the cell division cycle~\cite{Munzner2019}. Hence, component level formalisms are insufficiently detailed for precision medicine.




\textbf{3)}
The reaction-contingency based formalisms contain two layers of knowledge: the first (reactions) defines which states can change in the network, and the second (contingencies) defines how those reactions depend on previous state changes.
By describing both in terms of (and combinations of) site-specific states, the representation has the same granularity as the data, and the adaptive resolution necessary to be able to describe different degrees of knowledge about different reactions. The potential of this approach has been demonstrated with the comprehensive mechanistic model of cell division control in yeast \cite{Munzner2019}.

Currently, the main limitation of reaction-contingency based formalisms is the lack of automated reasoning tools. 
While microstate based models enjoy ample support of an extensive mathematical toolbox (\eg ODE solvers),
the same cannot be said about reaction-contingency models.
It is therefore that we regard the inclusion of KR techniques into the field of \sysbio, and specifically within reaction-contingency models, as a vital step in pursuing precision medicine -- especially as this is the only model class capable of modelling cellular signalling processes at the right resolution and at scale.


A fundamental problem that automated reasoning tools (adopted from KR) should address is the current lack of quantitative simulation methods for reaction-contingency models. 
Unfortunately, simply adapting techniques from microstate models will not be a viable option as this requires explicit enumeration of non-disjoint states.
Hence, novel approaches are needed to extend the simulation methods from qualitative Boolean models with a very limited representation of time into models that can capture quantitative and temporal aspects (across patients and treatments). Furthermore, these methods will need to be able to infer the state of latent variables from extremely sparse (temporal) data \cite{Rother2013}.

Given that reaction-contingency networks appear to be the only current method with the detail and scalability required to represent cellular signal transduction, as well as  its natural representation in logical formulas, we are convinced that KR with its automated reasoning tools has an important role to play in extending these capabilities reaction-contingency networks.

\section{A Glimpse into the Future}
\label{sec:marriage}

Given that biological knowledge is expressible in logic, in principle, we regard logic as the natural meeting ground for KR and \sysbio.
Note that the idea of modelling a regulatory network using Boolean representations has a rich history and dates back to \cite{Kauffman1969}.

The advantage with representing biological models using Boolean formulas is clear: they only require qualitative knowledge of the system and are easy to simulate.  
However, certain limitations exist as well.
For instance, purely deterministic propositional logic formulas do not allow for modelling stochasticity. Considering the noisy nature of cell biology this is a
rather important limitation.
Using logic as the assembly language for expressing biological knowledge, we can deploy well-studied techniques from the KR community to alleviate this issue. An obvious candidate are probabilistic or weighted logic rules.
We illustrate this in Figure~\ref{fig:rcgraph2poolnet}.
In the KR community it is well known that higher level languages such as Bayesian networks or probabilistic logic programs, which have already been applied in \sysbio~\cite{Woolf2005bayesian,Frohlich2015,Sachs2009learning,Gross2019}, can be compiled-down into weighed propositional logic formulas.
\footnote{We would like to underscore that we regard weighted propositional logic as an assembly language for expressing systems biology knowledge and that systems biology practitioners might prefer using higher level modelling languages, such as fragments of weighted first-order logic, to elicit their knowledge.}

Using probabilistic logics we are now able to model not only whether a reaction happens or not, but also what the probability is of this reaction happening within a certain time window. 
This increase in expressiveness comes, however at a computational cost.
For example, while simulating bipartite Boolean models in the deterministic case is a simple matter of recursively evaluating a Boolean function, this changes to performing probabilistic inference in temporal domains.
This changes the computational complexity from being simply linear in the size of the model to a \#P-hard problem \cite{valiant1979complexity,sang2005performing}.
Luckily, the field of KR has developed potent techniques to mitigate this hardness, \eg knowledge compilation~\cite{darwiche2002knowledge} and weighted model counting~\cite{darwiche2009modeling}.

Using existing biological knowledge as a scaffold for parametrizing stochastic functions allows us to incorporate valuable domain knowledge in models for precision medicine.
However, the question of picking the right set of parameters has not been answered.
We envisage to resolve this by opting for a learning approach over a modelling approach, for instance by using the well-established technique of first performing a knowledge compilation step and then performing gradient based optimization ~\cite{darwiche2003differential}.

This leads to the interesting observation that \textbf{\sysbio and KR are currently tackling an identical problem: integrating learning and reasoning.} 
Systems biologists approach the issue from an application-driven perspective (in the context of precision medicine), while computer scientists have an algorithmic perspective. 

\begin{figure}
    \centering
    \includegraphics[width=0.9\linewidth]{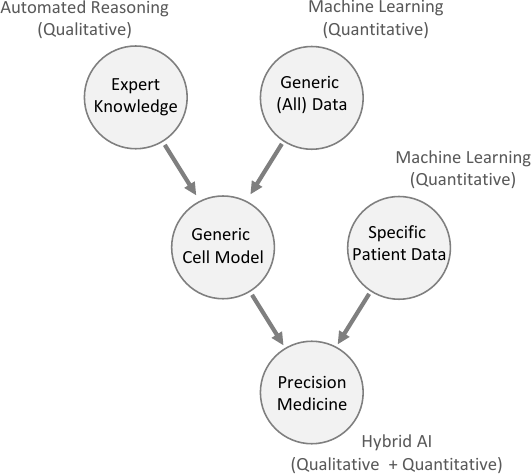}
    \caption{
    Schematic overview of our vision for achieving precision medicine: bio-medical background knowledge provides the structure for modelling signal transduction networks. This structure is then parametrized using generic (population level) data. Once this has been achieved, a generic model is refined on patient specific data. For example, via performing conditional probabilistic inference or fine-tuning parameters via gradient-based learning on patient specific data. 
    }
    \label{fig:circles2}
\end{figure}

Apart from tackling the important problem of solving precision medicine, this overlap of interests has also scientific merit.
On the one hand the field of KR can contribute existing learning algorithms from the probabilistic AI side ~\cite{Kitson2023,salam2021,choi2020probabilistic} and the emerging field of neuro-symbolic AI~\cite{MarraDMR24,Derkinderen2024semirings}.
On the other hand the field of systems biology can contribute an applications perspective, something that is occasionally missing in the KR research. This would allow for driving algorithmic advancements on a needs basis rather than a desires basis.
We give a diagrammatic overview of our vision for the technical integration of \sysbio and KR in the context of precision medicine in Figure~\ref{fig:circles2}.

\section{Conclusion}

Precision medicine requires a system level understanding, that is fundamentally grounded on lower level knowledge and observations.
As such, specific modelling choices and assumption have to be made in practice. 
For instance, the decision whether to use synchronous or asynchronous network simulations fundamentally affects the semantics of the modelling language~\cite{Garg2008,Schwab2020}. 
We have argued in this paper that these sort of modelling challenges ought to be addressed by a close collaboration between the KR and \sysbio communities where logic forms the common meeting ground.

This will require biologists to embrace to a wider extend the formalisms from computer science but it will also necessitate KR practitioners to make their tools more widely available and accessible, and to adapt them to the nature of the knowledge in the SysBio domain.
We believe that such an integration of KR techniques into the \sysbio toolbox is necessary to advance the field of precision medicine.







\section*{Acknowledgements}
This work was supported by the Wallenberg AI Autonomous Systems and Software Program (WASP) funded by the Knut and Alice Wallenberg Foundation, by the EU H2020 ICT48 project “TAILOR” under contract \#952215, and by the Exploring Inflammation in Health and Disease (X‐HiDE) Consortium, which is a strategic research profile at Örebro University supported by the Knowledge Foundation (20200017), and by strategic grants from Örebro University.




\bibliographystyle{kr}
\bibliography{references}

\clearpage
\newpage

\end{document}